\documentstyle[aps,twocolumn,epsf]{revtex}
\newcommand{\beq}{\begin{equation}}
\newcommand{\eeq}{\end{equation}}
\newcommand{\beqa}{\begin{eqnarray}}
\newcommand{\eeqa}{\end{eqnarray}}
\newcommand{\bM}{\bar{M}}
\newcommand{\bm}{\bar{m}}
\newcommand{\blam}{\bar{\lambda}}
\newcommand{\bV}{\bar{V}}
\renewcommand{\d}{\partial}
\newcommand{\btem}{\bibitem}

\begin{document}
%\draft
%%%%%%%%%%%%%%%%%%%% Title etc. %%%%%%%%%%%%%%%%%%

\title{Phase Structure of the Massive Scalar $\phi^4$ Model
        at Finite Temperature \\
   --- Resummation Procedure a la RG Improvement ---}

\author{Hisao Nakkagawa\cite{nakk} and Hiroshi Yokota\cite{yoko}}

\address{Institute for Natural Science, Nara University,
 1500 Misasagi-cho, Nara 631-8502, Japan}

\date{\today}

\maketitle

\begin{abstract}

In this paper a resummation method inspired by the renormalization-group improvement is applied to the one-loop effective potential (EP) in massive
scalar $\phi^4$ model at $T\neq0$.
By investigating the phase structure of the model at $T \neq 0$ we
get the following observations;
i) Starting from the perturbative
calculations with the theory renormalized at an arbitrary mass-scale
$\mu$ and at an arbitrary temperature $T_0$, we can in principle fully resum
terms of $O(\lambda T/\mu)$ together with terms of $O(\lambda (T/\mu)^2)$.
The key idea is to fix the arbitrary RS-parameters so as to make both of
one-loop radiative corrections to the mass as well as to the coupling vanish,
ensuring the function form of the EP to be determined up to the
next-to-leading order of large correction terms,
thus absorbing completely those terms of $O(\lambda (T/\mu)^2)$ and of
$O(\lambda T/\mu)$.
ii) If we start from the theory renormalized at the
temperature of the environment $T$, the $O(\lambda (T/\mu)^2)$-term resummation
can be simply completed through the $T$-renormalization itself.
With the lack of freedom we can set only one RS-fixing condition to absorb
the large terms of $O(\lambda T/\mu)$, thus only the partial resummation of
these terms can be carried out.
iii) In the above two analyses the temperature-dependent phase transition
of the model is shown with the analytic evaluation to proceed through the
second order transition.  The critical exponents are estimated analytically
and are compared with those in other analyses.
iv) Our resummation method does not work if we start from the theory
renormalized at $T=0$.
\end{abstract}

\pacs{11.10.Wx,12.38.Cy,12.38.Mh,11.30Rd}

\narrowtext

%%%%%%%%%%%%%% Text %%%%%%%%%%%%%%%%%%%%%% 

\section{Introduction and summary}

To investigate the phase structure of relativistic quantum field
theory, the effective potential (EP) is widely used as a powerful
and convenient tool\cite{Jac}. Perturbative calculation of the EP
especially at finite temperature in terms of the loop-wise expansion,
however, suffers from various troubles: the unreliability of the
perturbation theory\cite{ArEs}, the poor convergence or the breakdown of
the loop expansion\cite{Linde}, and also the strong dependence on the
artificially chosen renormalization-scheme (RS)\cite{Muta} such as the
renornalization scale $\mu$ and the renormalization
temperature $T_0$.
All these troubles have essentially the same origin, i.e., they come
about with the emergence of large perturbative correction terms
(large-{\it log} terms in vacuum theory, and large-$T$ ($T^2$) terms in
addition in the thermal theory) which depend explicitly on the RS.
Taking this fact into account, to break a way out of the above troubles
we must carry out the systematic resummation\cite{Braaten,DoJa,Lenpold}
of at least the dominant large correction terms, and at the same time
we must also solve the problem of the RS-dependence.
After the systematic resummation of higher order terms we can construct
effective field theories\cite{Braaten,Ian,Chiku,Nieto}, with which new
perturbation theory can be developed. Then we may have some hope that
such a resummation method can also work as a calculational procedure
for incorpolating the essential nonperturbative effect into the EP,
thus helping us to understand
the phase structure of the theory, toward which a variety of methods
has been
used\cite{ArEs,Braaten,DoJa,Lenpold,Ian,Chiku,Nieto,AmPi,Ameli,ArYa,Roh,Ogure,BB,eps}.

Recently simple but very efficient renormalization group (RG) improvement
procedures for resumming dominant large correction terms are
proposed in vacuum\cite{Bando} and in thermal\cite{NY1} field theories.
This procedure was originally proposed to solve the problem of the
strong RS-dependence of the EP calculated through the loop-expansion
method. After the RG improvement, the RG improved perturbation theory
can be consistently formulated, thus we may also have hope that this can
serve as a tool for incorpolating the nonperturbative effect into the EP.
Application of this RS improvement procedure to the $O(N) \phi^4$ model
at large-$N$ has revealed\cite{NY1} that it also carry out the systematic
resummation of large correction terms to all orders, showing the exact
second order nature of the temperature-dependent phase transition of the
model.
 
In this paper we briefly explain this resummation procedure a la RG,
namely the RG-improvement procedure, and present the results of application
of this procedure to the massive scalar $\phi^4$ model at finite
temperature, showing that our RG-improvement procedure not only
resolves the problem of the RS-dependence, but also incorpolates
important non-perturbative effects.

It is worth noticing here that in the massive scalar $\lambda \phi^4$ model
at high temperature the large correction terms appearing in the $L$-loop
EP have the structures as follows; i) terms proportional to powers of the
temperature $T$: i-a) $(\lambda (T/\mu)^2)^L$, i-b) $(\lambda T/\mu)^L$, both
up to powers of $\ln (T/\mu)$, and i-c) $(\lambda \ln (T/\mu))^L$, 
and ii) terms proportional to
powers of the ordinary {\it log}\,s, $(\lambda \ln ({\cal M}/\mu))^L$, 
where ${\cal M}$ is
the large mass scale appearing in the theory. To investigate the phase
structure of the model we must effectively resum at least the
``dominant'' perturbative correction terms systematically to all orders.

We have found that the proposed resummation procedure a la RG works
efficiently in the massive scalar $\phi^4$ models, not only by resolving the
problem of strong RS-dependence, but also by properly as well as
systematically resumming terms having the structures i-a) and i-b) above.
It is worth mentioning that our procedure is essentially the RG-improvement,
which enables us to formulate the RG-improved effective perturbation theory
without any trouble such as, e.g., the double-counting of higher loop
diagrams.
It is also to be noted that, except in the $O(N)$ symmetric model in
the large-$N$ limit, in order for the present resummation mothod to work
the starting perturbative calculation should be
performed with the theory renormalized at non-zero renormalization
temperature $T_0 \neq 0$~\cite{NY2}.

Main outcomes of the present anaysis are the followings;

The massive scalar $\phi^4$ models, irrespective of the number of
components of the scalar field $\phi$, have an ordinary phase in which
the effective potential changes its form as the temperature increases from
the symmetry-broken wine-bottle form at low temperature to the
symmetry-restored one at high temperature. It should be stressed that the
temperature dependent phase transition of the massive scalar $\phi^4$
models is shown to proceed {\it through the second order transition}.
The critical exponents estimated analytically are compared with those
in other analyses, showing
siginificant deviations from the mean-field values and the reasonable
agreement with ``experiments''\cite{experi,lattice}. 

In the simple single component
model below the critical temperature $T_c$, in addition to
the ordinary symmetry-broken phase there appears a new phase
with the potential unbounded from below. This two-phase structure at
low temperature survives in the zero temperature limit, indicating the
simple $\phi^4$ model being an unstable theory. The $O(N)$ symmetric
model in the large-$N$ limit exists as a stable theory without
having such an unstable phase.

These results are obtained by performing the systematic resummation of
terms of $O(\lambda (T/\mu))$ through the RG-improvement of the one-loop
calculation. Further anayses including the more precise evaluation of
critical exponents with numerical anaysis as well as the improvement of
two-loop calcualtion will be given elsewhere\cite{NY3}.

\section{Resummation a la RG Improvement}

Let us focus on the massive self-coupled scalar $\phi^4$ model
at finite temperature,
\beq
  {\cal L} = \frac{1}{2} (\d_{\mu} \phi )^2 - \frac{1}{2} m^2 \phi^2
      - \frac{1}{4!} \lambda \phi^4 - h m^4 \ \ \ (m^2 < 0) \ ,
\eeq
and renormalize the theory at an arbitrary mass-scale $\mu$ and
at an arbitrary renormalization-temperature $T_0$.
Note that we now have at least two arbitrary parameters (scales)
that specify the present RS (hereafter we call this scheme as the
$T_0$-renormalization). Then the key idea to resolve the
RS-ambiguity is to use correctly and efficiently the fact that
the exact EP satisfies a set of renormalization group equations
(RGE's)\cite{Matsu} with respect to changes of the arbitrary parameters
$\mu \to \bar{\mu}=\mu e^{t}$ and $T_0 \to \bar{T}_0 = T_0 e^{\rho}$,
\beqa
    \left(\frac{\d}{\d t} +
        \beta_t \frac{\d}{\d \lambda} -
        m^2 \theta_t \frac{\d}{\d m^2} -
        \phi \gamma_t \frac{\d}{\d \phi} + 
        \beta_{ht} \frac{\d}{\d h} \right) &V&  \nonumber \\
                          = &0&   \ , \\
    \left( \frac{\d}{\d \rho} +
        \beta_{\rho} \frac{\d}{\d \lambda} -
        m^2 \theta_{\rho} \frac{\d}{\d m^2} -
        \phi \gamma_{\rho} \frac{\d}{\d \phi} +
        \beta_{h\rho} \frac{\d}{\d h} \right) &V&  \nonumber \\
                          = &0&  \ ,
\eeqa
where a scaled variable $\xi \equiv T_0/\mu$ is introduced.
Solution to the RGE's is
($\bar{\mu}^2 \equiv \mu^2 e^{2t}, \bar{\xi}^2 \equiv \xi^2 e^{2\rho}$)
\beqa
    & & V \left( \phi, \lambda, m^2, h, T; \mu^2, \xi^2 \right) = \nonumber \\
    & & \bV \left( \bar{\phi}(t, \rho), \blam(t, \rho), \bm^2(t, \rho),
        \bar{h}(t, \rho), T; \bar{\mu}^2, \bar{\xi}^2 \right), 
\eeqa
where the barred quantities $\blam$, $\bm^2$, etc. are the RG-improved running
parameters whose responces to the changes of $t$ and $\rho$ are determined by
the coefficient functions of the RGE's, $\beta$'s,
$\theta$'s etc., 
\beq
  \beta_x = \frac{\d \lambda}{\d x}, \ \ 
  \theta_x = - \frac{\d \ln m^2}{\d x}, \ \ \mbox{etc.},
  \ \ \mbox{with} \ \ x=t \ \ \mbox{or} \ \ \rho \ ,
\eeq
with the boundary condition that the barred quantities are reduced to the
unbarred paprameters at $t = \rho=0$.
Thus, {\it the EP is completely determined once its function form is known at
certain values of $t$ and $\rho$}. The problem of resolving the RS-dependence
of the EP is now reduced to the one {\it how can we determine, with the limited
knowledge of the L-loop calculation, the function form of the EP}.

When we renormalize the theory at some definite values of $T_0$, then the RGE's as
well as the corresponding responce-equations with respect to the change of
$\rho$ in the above Eqs.(2)-(5) do not appear at all.
Two choices $T_0 = 0$ and $T_0 =T$  ($T$ : the temperature of the environment)
are of interest, and hereafter we call them as the $T = 0$ renormalization and
the $T$-renormalization respectively.  In these two RS's we only need to study
the responces with respect to the change of surviving one arbitrary parameter
$t$.

Let us notice here that in the scalar $\phi^4$ model (at least in the $O(N)$ symmetric model in $N \to \infty$) the dominant large corrections appear as
a power function of the effective variable $\tau$
\beqa
 & & \tau/\lambda \equiv \Delta_1 \\
 & & \begin{array}{lll} 
      = & \displaystyle{
            \frac{T^2}{2 \pi^2M^2} L_1 \left( \frac{T^2}{M^2} \right)
            - \frac{T_0^2}{24M^2} } & \\ 
       & & \\
         & \displaystyle{- \frac{1}{2 \pi^2} L_2(\xi^2)
            + \frac{b}{2} \left( \ln \frac{M^2}{\mu^2} -1 \right)} &
               \mbox{for} \ T_0 \\ 
       & & \\
      = & \displaystyle{
              \frac{T^2}{2 \pi^2M^2} \left\{ L_1 \left( \frac{T^2}{M^2} \right)
             - \frac{\pi^2}{12} \right\} } & \\
       & & \\
        & \displaystyle{- \frac{1}{2 \pi^2} L_2 \left(
               \frac{T^2}{\mu^2} \right)
               + \frac{b}{2} \left( \ln \frac{M^2}{\mu^2} -1 \right)} &
                \mbox{for} \ T  \\
       & & \\
      =&  \displaystyle{
             \frac{T^2}{2 \pi^2M^2} L_1 \left( \frac{T^2}{M^2} \right) 
               + \frac{b}{2} \left( \ln \frac{M^2}{\mu^2} -1 \right)} &
                \mbox{for} \ T=0 
    \end{array} \nonumber
\eeqa
where $b=1/16\pi^2$, $M^2 = m^2 + \lambda \phi^2/2$ and
\beq
 M^2 \Delta_1 \equiv \frac12 \sum\!\!\!\!\!\!\!\int \frac{1}{k^2-M^2} +
       \mbox{(one-loop counter term)}, 
\eeq
\beqa
 L_1 \left( \frac{1}{a^2} \right) &\equiv& \frac12
       \int_0^{\infty} \frac{k^2 \, dk}{\sqrt{k^2+a^2}} \frac{1}{
         \exp (\sqrt{k^2+a^2}) - 1}, \\
 L_2 \left( \frac{1}{a^2} \right) &\equiv& - \frac14
       \int_0^{\infty} \frac{dk}{\sqrt{k^2+a^2}} \frac{1}{
         \exp (\sqrt{k^2+a^2}) - 1}.
\eeqa
$M^2 \Delta_1$ is nothing but (a part of) the renormalized one-loop self-energy
correction, Fig. 1a, having the high temperature behavior (exact up to $T$-independent
constant)
\beqa
 & & \tau/\lambda \\
 & & \begin{array}{lll} \sim & \displaystyle{\frac{T^2-T_0^2}{24M^2}
                       - \frac{T}{8\pi M} + \frac{T_0}{16\pi \mu}
                       + b \ln \frac{T}{T_0}} & \quad \mbox{for $T_0$} \\
            \sim & \displaystyle{- \frac{T}{8\pi M} + \frac{T}{16\pi \mu}}
                           & \quad \mbox{for $T$} \\
            \sim & \displaystyle{\frac{T^2}{24M^2}
                       - \frac{T}{8\pi M} + b \left( \ln \frac{4\pi T}{\mu} 
                       - \gamma_E \right) }
                           & \quad \mbox{for $T=0$}.
 \end{array} \nonumber
\eeqa

%%%%%%%%%%%%%%%%%%%%%%%%%%%%%%%%%%%%%%%%%%%%%%%%%%%%%%%%%%%%%%%%%%%%
\begin{figure}\epsfxsize=4.5cm
\centerline{\epsfbox{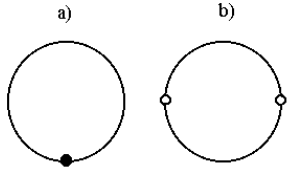}}
\vskip4mm
\caption{The one-loop diagrams: a) with one vertex corresponding
      to $\Delta_1$ , and b) with two vertices corresponding to $\Delta_2$ .}
\label{fig1}
\end{figure}
%%%%%%%%%%%%%%%%%%%%%%%%%%%%%%%%%%%%%%%%%%%%%%%%%%%%%%%%%%%%%%%%%%%

Since in the $O(N)$ symmetric model in the large-$N$ limit the EP can be
expressed\cite{NY1} in the power-series expansion in $\tau$;
\beq
    V =  \frac{M^4}{\lambda}
     \sum_{\ell =0}^{\infty} \lambda^{\ell} \: \left[ \: 
               F_{\ell}(\tau) + z \delta_{\ell ,0}
            \:  \right] \ , \ \ \ z \equiv \frac{\lambda h m^4}{M^4} ,
\eeq
where
\beqa
     F_{\ell}(\tau) \equiv  \sum_{L=\ell}^{\infty} v_{\ell}^{(L)}
             \tau^{L- \ell} \ ,& & \ \ F_{0}(\tau) = \sum_{L=0}^{2} v_{0}^{(L)} \tau^{L}\ , \nonumber \\
             & & v_0^{(L)}=0 \mbox{ for } L \ge 3 \ ,
\eeqa
then we can easily find the solution\cite{NY1} to the above posed problem;
\underline{at $\tau=0$}, the ``$\ell$th-to-leading $\tau$'' function $F_{\ell}$
is given solely in terms of the $\ell$-loop level potential,
$F_{\ell}(\tau=0)=v_{\ell}^{(L=\ell)}$, {\it which is a pure constant}.
So if we caluculated the EP to the L-loop level, then at $\tau=0$ it already
gives the function form ``exact'' up to ``$L$th-to-leading $\tau$'' order.
With the $L$-loop potential at hand, the EP satisfying the RGE's can be given by
\beqa
     V &=& \bM^4(t)  \left.
      \sum_{\ell =0}^{L} \blam^{\ell -1}(t)
       \left[ \: \bar{v}_{\ell}^{(\ell)}(t) +
            \bar{z}(t) \delta_{\ell,0} \:
            \right] \right|_{\bar{\tau}(t)=0}     \nonumber  \\
   &=&  V_L  \left. ( \bar{\phi}(t), \blam(t),
             \bm^2(t), \bar{h}(t);
            \mu^2 e^{2t}) \right|_{\bar{\tau}(t)=0} \ ,
\eeqa
where the barred quantities should be evaluated at such a $t$ satisfying
$\tau(t)=0$. 

In the simple $\phi^4$ model (and also in the general $O(N)$ symmetric model), however, because of the existence of the three-point coupling the structure of
the EP in its perturbation series expansion
is not the simple power-series expansion
in $\tau$, Eqs.(11) and (12). Thus we are forced to study more carefully the
sub-structure of the perturbation series expression of the EP. 
We find that there is an another important effective variable $\kappa$
\beqa
 & & \kappa/\lambda \equiv \Delta_2 \\
 & & \begin{array}{lll} 
      = & \displaystyle{
            \frac{1}{2 \pi^2}\left\{ L_2 \left( \frac{T^2}{M^2} \right)
            - L_2 (\xi^2) \right\} } & \\ 
       & & \\
         & \displaystyle{+ \frac{b}{2}\ln \frac{M^2}{\mu^2}} &
               \mbox{for} \ T_0 \\ 
       & & \\
      = & \displaystyle{
              \frac{1}{2 \pi^2}\left\{ L_2 \left( \frac{T^2}{M^2} \right)
            - L_2 \left( \frac{T^2}{\mu^2} \right) \right\} } & \\
       & & \\
        & \displaystyle{
               + \frac{b}{2}\ln \frac{M^2}{\mu^2} } &
                \mbox{for} \ T  \\
       & & \\
      =&  \displaystyle{
             \frac{1}{2 \pi^2} L_2 \left( \frac{T^2}{M^2} \right) 
              + \frac{b}{2} \ln \frac{M^2}{\mu^2} } &
                \mbox{for} \ T=0 
    \end{array} \nonumber
\eeqa
where
\beq
 \Delta_2 \equiv \frac12 \sum\!\!\!\!\!\!\!\int \frac{1}{(k^2-M^2)^2} +
       \mbox{(one-loop counter term)},
\eeq
which is nothing but the renormalized one-loop correction to the coupling,
Fig.1b, having the high temperature behavior (exact up to $T$-independent
constant)
\beqa
 & &  \kappa/\lambda \\
 & & \begin{array}{lll} \sim & \displaystyle{\frac{T_0}{16\pi \mu} -
                      \frac{T}{16\pi M} 
                       + b \ln \frac{T}{T_0}} & \quad \mbox{for $T_0$} \\
            \sim & \displaystyle{\frac{T}{16\pi \mu} - \frac{T}{16\pi M}}
                           & \quad \mbox{for $T$} \\
            \sim & \displaystyle{ - \frac{T}{16\pi M} 
                      + b \left( \ln \frac{4\pi T}{\mu} 
                       - \gamma_E \right) }
                           & \quad \mbox{for $T=0$}.
 \end{array} \nonumber
\eeqa
$\Delta_2$ represents a diagram that appears as an important sub-diagram
producing large temperature-dependent corrections in the higher loop diagramms.  

With the two effective variables $\tau$ and $\kappa$ in hand, the EP in the
massive scalar $\phi^4$ model can be expressed as a double power-series
expansion in these two variables; Eq.~(11) where $F_{\ell}$ now having
the double power-series expression in $\tau$ and $\kappa$,
\beq
   F_{\ell} (\tau , \kappa ) = \sum^{\infty}_{L=\ell} \ 
           \sum_{(L_1,L_2; \ell_1,\ell_2)} 
           v^{(L;L_1,L_2)}_{(\ell;\ell_1,\ell_2)} \,
           \tau^{L_1 - \ell_1} \, \kappa^{L_2 - \ell_2} \ .
\eeq
The summation over $(L_1,L_2; \ell_1,\ell_2)$ should be
taken with the constants $L_1+L_2=L$, $ \ell_1 + \ell_2 = \ell$ 
and $L_i \ge \ell_i \ (i=1,2)$.

Then the solution to the problem posed above,
namely the problem how can we determine, with the limited knowledge of the
$L$-loop calculation, the function form of the EP, can be found as a simple generalization of the solution in the $O(N)$ symmetric model;
{\it at $\tau=\kappa=0$, the ``$\ell$th-to-leading'' function $F_{\ell}$
is given solely in terms of the $\ell$-loop level potential,
$F_{\ell}(\tau=\kappa=0)=v_{\ell}^{(L=\ell)}$},
where $v^{(L=\ell)}_{\ell} \equiv v^{(L; L_1=\ell_1,L_2=\ell_2)}_{(\ell; \ell_1,
\ell_2)}$.
Note that in the present case, being different from the large-$N$ limit of the
$O(N)$ symmetric model, $v_{\ell}^{(L=\ell)}$
can be a function of $\ln (M/\mu )$, $\mu/T$,
and of $M/T$,
but is importantly {\it less than $O((T/\mu )^{\ell-1})$.
So if we caluculated the EP to the $L$-loop level, then
at $\tau=\kappa=0$
it already gives the function form ``exact'' up to ``$L$th-to-leading'' order}.
With the $L$-loop potential at hand, the EP satisfying the RGE's can be given by
\beqa
     V &=& \bM^4
      \sum_{\ell =0}^{L} \blam^{\ell -1} \left.
       \left[ \: \bar{v}_{\ell}^{(\ell)} +
            \bar{z} \delta_{\ell,0} \:
            \right] \right|_{\bar{\tau}=\bar{\kappa}=0} \nonumber \\
       &=& \left. V_L (\bar{\phi}, \blam, \bm^2, \bar{h}, T; \bar{\mu}^2, 
              \bar{\xi}^2 ) \right|_{\bar{\tau}=\bar{\kappa}=0} \ ,
\eeqa
where the barred quantities, which are functions of $t$ and $\rho$,
should be evaluated at such $t$ and $\rho$ satisfying
\beq
      \bar{\tau}(t,\rho)=\bar{\kappa}(t,\rho)=0.
\eeq

In the RG-improved EP, Eq.~(18), the highest series of contributions take the
form $\blam (\blam T/\bar{\mu})^{\ell -1}$, namely, contributions of
$O \left( (\lambda T/\mu)^{\ell} \right)$ nor of 
$O \left( (\lambda T^2/\mu^2)^{\ell} \right)$ never appear explicitly,
all resummed and absorbed into the barred quantities.

It should be noted that the two independent conditions, Eqs.(19), now fix the
RS, namely guarantee to carry out the RG-improvement of the EP in the sense
noted above. These two conditions actually make the one-loop radiative
corrections to the mass and
to the coupling totally vanish. Generally speaking we can find solutions
$t$ and $\rho$ to Eqs.(19) if the starting perturbative calculation is
performed in the $T_0$-renomalization, otherwise not.     

\section{Phase structure of the simple massive scalar $\phi^4$ model at $T\neq0$}

Now we explicitly apply the RG improvement procedure explained above to the
massive scalar $\phi^4$ model at $T\neq0$, and study the phase structure.
Here we only carry out the improvement of the one-loop results. Thorough
analysis including those of critical exponents and
the improvement of the two-loop results will be given
elsewhere\cite{NY3}.

\subsection{$T_0$-renomalization}

The perturbatively calculated one-loop EP in the $T_0$-renomalization is
\beqa
 V_1 &=& \frac12 m^2 \phi^2 + \frac{1}{4!} \lambda \phi^4 + hm^4 \nonumber \\
     & & + \frac{M^4}{2} \left[ \frac{b}{2} \left( \ln \frac{M^2}{\mu^2} -
            \frac32 \right) + \frac{T^4}{\pi^2 M^4} L_0
            \left( \frac{T^2}{M^2} \right) \right. \nonumber \\
     & & \ \ \ \ \  \left. - \frac{T_0^2}{12 M^2} - \frac{1}{2\pi^2} L_2 (\xi^2)
            \right] \ ,
\eeqa
where
\beq
    L_0 \left( \frac{1}{a^2} \right) \equiv
       \int_0^{\infty} k^2 \, dk \, \ln [ 1 - \exp \{ - \sqrt{k^2+a^2} \} ]
       \ .
\eeq
 
At the one-loop level the mass-squared $m^2$ and the coupling $\lambda$
satisfy the RGE's,
\beqa
& & \frac{\d \lambda}{\d t} = \beta_t = 3 b \lambda^2, \ \ \ 
 \frac{\d \lambda}{\d \rho} = \beta_{\rho} = 
        \frac{3}{2 \pi^2} \left( \frac{d}{d \rho} L_2(\xi^2) \right) \lambda^2, \\
& & \frac{\d \ln m^2}{\d t}= -\theta_t = - b \lambda, \ \nonumber \\
 & & \frac{\d \ln m^2}{\d \rho} = - \theta_{\rho} = 
        - \left\{ \frac{1}{2 \pi^2} \frac{d}{d \rho} L_2(\xi^2)
        + \frac{T_0^2}{12 m^2} \right\} \lambda \ .
\eeqa
To perform the RG improvement explained in the last Sec.~II, we must solve
these RGE's to obtain $\blam$ and $\bm^2$.
Responce-equations of $\lambda$, Eqs.(22), can be solved analytically to get 
\beqa
  \blam &=& \lambda F^{-1} (\bar{\mu}, \bar{\xi}) , \\
  F (\bar{\mu}, \bar{\xi}) &=& 1 - 3 \lambda \left[
          b \ln \frac{\bar{\mu}}{\mu} + \frac{1}{2\pi^2} \left\{
               L_2 (\bar{\xi}^2) - L_2 (\xi^2) \right\} \right] \ .
\eeqa
The differential equations (23) describing the responces of $m^2$ can not be
solved exactly even at the one-loop level in the general $T_0$-renomalization.
Thus to study the definite results of the one-loop improvement, we must perform
the numerical integration of Eqs.(23). Fortunately, however, we can find an
approximate solution of $\bm^2$ in the high temperature regime,
\beqa
  \bm^2 &=& \hat{m}^2 F^{-1/3} (\bar{\mu}, \bar{\xi}) , \\
  \hat{m}^2 &=& m^2+\frac{1}{24} \lambda \bar{T}_0^2\ .
\eeqa
The solution $\bm^2$ is exact up to order $T/\mu$ in the high temperature
(HT) expansion.

Up to now $\bar{\mu}$ and $\bar{\xi}$ in the above eqations (24)-(27) can be arbitrary, with $\mu$ and $\xi$ being fixed at the initial values of
renormalization. {\it Our RG-improvement procedure, i.e., the resummation
procedure
a la RG, can then be carried out by choosing the RS-fixing parameters
$\bar{\mu}$ and $\bar{\xi}$ so as to satisfy $\bar{\tau}(t,\rho)=\bar{\kappa}(t,\rho)=0$, Eqs.(19), namely to make the
one-loop radiative corrections to the mass and also to the coupling fully
vanish.} First let us see the solutions to the above RS-fixing equations.
In the HT regime where $T/\mu \gg 1$, we can use the HT expansion of the
functions  $L_1(a^{-2})$ and $L_2(a^{-2})$ with $a^{-1}=T/\mu$,
\beqa
   L_1(a^{-2}) &\simeq& \frac{\pi^2}{12} - \frac{\pi}{4} a
                 - \frac18 a^2 \ln a  \nonumber \\
             & & \ \ - \frac18 \left( \gamma_E - \ln 4\pi - \frac12 \right) a^2, \\
   L_2(a^{-2}) &\simeq& - \frac{\pi}{8a} - \frac18 \ln a
                 - \frac18 (\gamma_E - \ln 4 \pi) .
\eeqa
Then the RS-fixing equations $\bar{\tau}(t,\rho)=\bar{\kappa}(t,\rho)=0$ give two
equations being exact up to $T$-independent constant,
\beqa
     & &  \left( \frac{T}{\bM} \right)^2 - \left( \frac{\bar{T}_0}{\bM} \right)^2 
           - 24\pi b \frac{T}{\bM} = 0 \ , \\
     & & \frac12 \ln \frac{T}{\bar{T}_0} + \pi \left( \frac{\bar{T}_0}{\bar{\mu}}
           - \frac{T}{\bM} \right) = 0 \ ,
\eeqa
where $\bM^2 = \bm^2 + (1/2)\blam \phi^2$. Solutions to these equations (30) and
(31) are
\beqa
    \bar{T}_0^2 &=& T^2 \left( 1- \frac{3}{2\pi} \frac{\bM}{T} \right) \ , \\
    \bar{\mu} &=& \bM \left(1 - \frac{3}{4\pi} \frac{\bM}{T} \right) \ ,
\eeqa                                                                              determining the RS-parameters with which the EP should be evaluated.

If we can find the exact solution $\bm^2$ to the coupled equations (23)
together with the exact solution $\blam$, Eqs.(24) and (25), then with
the use of above solutions to the RS-parameters, Eqs.(32) and (33), we can
perform the full resummation of $O(\lambda T/\mu)$ terms together with terms
of $O(\lambda (T/\mu)^2)$. {\it Question}: How has the resummation of terms of
$O(\lambda (T/\mu)^2)$ been carried out?  {\it Answer}: It is done through the
$T_0$-renormalization with the RS-parameters (32) and (33),
guaranteeing the renomalized
mass-squared to have $T$-dependent term $\lambda T^2$. As explained above,
however, the solution $\bm^2$ with compact expression, Eqs.(26) and (27),
is an approximate one, thus may weaken the ``full resummation'' of
$O(\lambda /\mu)$ terms.

Now let us study the consequences of the RG-improvement in the
$T_0$-renormalization, with solutions $\blam$, Eqs.(24) and (25),
$\bm^2$, Eqs.~(26) and (27), $\bar{T}_0$ and $\bar{\mu}$,
Eqs.~(32) and (33).
The mass-gap equation,
\beq
   M^2 = \hat{m}^2 [ 1 - F^{2/3} (\bar{\mu}, \bar{\xi}) ]
      + \bM^2 F(\bar{\mu}, \bar{\xi})  \ ,       
\eeq
becomes at HT (up to $O(\lambda \ln (T/\mu))$)
\beq
   M^2 = \bM^2 + b \lambda ( 3 \bM^2 - 2 \hat{m}^2) \left[ \pi \frac{T}{\bM}
             - \ln \frac{T}{\bar{\mu}} \right] \ .
\eeq
With the mass-gap equation (35) we can study $\phi^2$ as a function of
$\bM^2$, Figs.2a-2c, from which we can see the followings; i) at
sufficient high temperature there is only one phase in the HT
approximation and $\bM^2$ can not reach zero, Fig.2c, ii) at the
critical temperature $T_C \sim \sqrt{24|m^2|/\lambda}\ $ the $\bM^2$
vanish at $\phi^2=0$, Fig.2b, and iii) below the critical temperature $T_C$
there appears a new phase in the small-$\bM^2$ region, thus the two-phase
structure comes about at low temperature, Fig.2a.

%%%%%%%%%%%%%%%%%%%%%%%%%%%%%%%%%%%%%%%%%%%%%%%%%%%%%%%%%%%%%%%%%%%%
%\begin{figure}\epsfxsize=8.5cm
%\centerline{\epsfbox{fig2.eps}}
%\vskip4mm
%\caption{The $\phi^2$-$\bM^2$ relation of the massive $\phi^4$ model
%      in the $T_0$-renormalization: a) $\tilde{T}_1=21.8$ ,
%      b) $\tilde{T}_2=\tilde{T}_C=21.893$, and c) $\tilde{T}_3=22.0$
%      with $\tilde{T} \equiv T/|m_0|$. $m_0$ is
%      the renormalized mass in the vacuum theory.}
%\label{fig2}
%\end{figure}
%%%%%%%%%%%%%%%%%%%%%%%%%%%%%%%%%%%%%%%%%%%%%%%%%%%%%%%%%%%%%%%%%%%

By studying the structures of the RG-improved one-loop EP, $\bV_1$,
\beqa
 \bV_1 &=& \frac12 \bm^2 \phi^2 + \frac{1}{4!} \blam \phi^4 + \bar{h}\bm^4
                        \nonumber \\
        & & + \frac{\bM^4}{2} \left[ - \frac{b}{4}
            + \frac{T^4}{\pi^2 \bM^4} L_0
            \left( \frac{T^2}{\bM^2} \right)  \right. \nonumber \\
       & & \left.
         - \frac{T^2}{2\pi^2 \bM^2} L_1 \left( \frac{T^2}{\bM^2} \right)
            - \frac{\bar{T}_0^2}{24 \bM^2} \right]  \\
       &=& \frac12 \bm^2 \phi^2 + \frac{1}{4!} \blam \phi^4-\frac{\bm^4}{2 \blam}
                        \nonumber \\
       & & + \frac{(T^2-\bar{T}_0^2) \bM^2}{48} - \frac{T \bM^3}{48 \pi} + \cdots ,
\eeqa
at the corresponding temperatures and phases, we can see the nature of the
temperature-dependent phase-transition of the model, Figs.3a-3c; i) At low
temperature below $T_C$ the EP has twofold structure representing the
existence of two phases, Fig.3a, the ordinary mass phase and the small mass
phase. The ordinary mass phase, with its counterpart in the tree-level
potential, has the EP with the wine-bottle structure with the minimum at
$\phi=\phi_0 \sim (T_C^4 |m^2|^3/\lambda)^{1/10}$.
The small mass phase is a new phase appearing as a result of resummation a la
RG, without having any counterpart in the tree-level potential. This new phase
is a ``symmetric'' phase with a linearly decreasing potential, namely with
a potential unbounded from below, indicating the simple $\phi^4$ model
becoming an unstable theory at low temperature. As the temperature becomes
higher the minimum of the ordinary mass phase eventually diminishes, and
ii) at the critical temperature $T_C$ the minimum of the potential at
non-zero $\phi$ completely disappears. The EP shows a symmetric
structure in $\phi$ with the minimum at $\phi=0$, $V(\phi)-V(0) \propto
\phi^{\delta+1}$, $\delta \sim 5.0$, Fig.3b, and iii) at high temperature
above $T_C$ the EP remains symmetric in $\phi$ whose positive
curvature near $\phi=0$ becomes strong as the temperature, Fig.3c.
Transition between the ordinary mass broken phase at low temperature and
the symmetric phase at high temperature clearly proceeds through the second
order transition.

%%%%%%%%%%%%%%%%%%%%%%%%%%%%%%%%%%%%%%%%%%%%%%%%%%%%%%%%%%%%%%%%%%%%
%\begin{figure}\epsfxsize=8.5cm
%\centerline{\epsfbox{fig3.eps}}
%\vskip4mm
%\caption{The RG improved effective potential of the massive $\phi^4$
%      model at three temperatures in the $T_0$-renormalization: 
%      a) $\tilde{T}_1=21.8$, b) $\tilde{T}_2=\tilde{T}_C=21.893$, and
%      c) $\tilde{T}_3=22.0$  with $\tilde{T} \equiv T/|m_0|$.
%      $\tilde{V} \equiv [ \bV_1 (\tilde{\phi}) - \bV_1 (\tilde{\phi}_{min}) ]/
%      |m_0|^4$, $\tilde{\phi} \equiv \phi/|m_0|$ and $\lambda = 1/20$.}
%\label{fig3}
%\end{figure}
%%%%%%%%%%%%%%%%%%%%%%%%%%%%%%%%%%%%%%%%%%%%%%%%%%%%%%%%%%%%%%%%%%%

To get more definite conclusion of the complete resummation of
$O(\lambda T/\mu)$, we need the numerical analysis in solving $\bm^2$ exactly
without HT approximation.

\subsection{$T$-renormalization}

The  perturbatively calculated one-loop EP in the $T$-renormalization is
\beqa
 V_1 &=& \frac12 m^2 \phi^2 + \frac{1}{4!} \lambda \phi^4 + hm^4 \nonumber \\
     & & + \frac{M^4}{2} \left[ \frac{b}{2} \left( \ln \frac{M^2}{\mu^2} -
            \frac32 \right) + \frac{T^4}{\pi^2 M^4} L_0
            \left( \frac{T^2}{M^2} \right) \right. \nonumber \\
     & & \ \ \ \ \  \left. - \frac{T^2}{12 M^2} - \frac{1}{2\pi^2}
            L_2 \left(\frac{T^2}{\mu^2}\right)
            \right] \ .
\eeqa
It is worth noting that in the $T$-renormalization the resummation of the
dominant $O(\lambda (T/\mu)^2)$ terms can be automatically performed
through renormalization, giving the renormalized mass-squared
\beq
  m^2 \simeq m_0^2 + \frac{\lambda T^2}{24} ,
\eeq
$m_0$ the renormalized mass in the vacuum theory.
$m^2$ appears as a mass-term in the propagator with which the perturbative
calculation is performed.

At the one-loop level we can find the exact
solutions of the running parameters $\bm^2$ and $\blam$,
\beqa
 & & \bM^2 = \bm^2 + \frac12 \blam \phi^2 , \ \ \bm^2 = m^2 f^{-1/3} , \ \
 \blam = \lambda f^{-1} ,  \\
 & & f = 1 - 3 \lambda \left[ bt + \frac{1}{2 \pi^2} \left\{ L_2 
             \left(\frac{T^2}{\bar{\mu}^2}\right) - L_2
             \left(\frac{T^2}{\mu^2}\right) \right\} \right].
\eeqa
As explained in the Sec.II, in the $T$-renormalization we have only one
arbitrary parameter $\mu$, and
we can set only one RS-fixing condition. Reminding us of the
lessons from previous works\cite{DoJa,AmPi,Ameli,NY1}, we choose
$\bar{\tau}(t, \rho=0)=0$ as a condition to fix the RS, namely choose
the RS-parameter $t$ so that the one-loop radiative corrections to
the mass vanish. Then the RG improvement can be
performed analytically, obtaining the improved EP as
\beqa
 \bV_1 &=& \frac12 \bm^2 \phi^2 + \frac{1}{4!} \blam \phi^4 + \bar{h}\bm^4
             \nonumber \\
       & &   + \frac{\bM^4}{2} \left[ -\frac{b}{4}
             + \frac{T^4}{\pi^2 \bM^4} L_0 \left( \frac{T^2}{\bM^2} \right) 
             \right. \nonumber \\
      & &   \left. - \frac{T^2}{2\pi^2 \bM^2} L_1 \left( \frac{T^2}{\bM^2} \right)
            - \frac{T^2}{24 \bM^2} \right] \\ 
      &=& \frac12 \bm^2 \phi^2 + \frac{1}{4!} \blam \phi^4 - \frac{\bm^4}{2 \blam}
             - \frac{T \bM^3}{48 \pi} + \cdots.
\eeqa
All the barred quantities should be evaluated at such a $t$ satisfying the
RS-fixing condition $\bar{\tau}(t, \rho=0)=0$, which gives the mass
gap equation
\beqa
   M^2 &=& m^2 + f(\bM^2)\bM^2 - f(\bM^2)^{2/3} m^2  \ , \\
   f(\bM^2) &\simeq& 1 + 3 \lambda \left[ \frac{T}{8 \pi \bM} - 
          \frac{T}{16 \pi \mu} \right] + \cdots.
\eeqa 

It should be noted that the RS-fixing condition $\bar{\tau}(t,\rho=0)=0$
gives the equation at HT
\beq
     \frac{T}{8 \pi \bM} - \frac{T}{16 \pi \bar{\mu}} = 0, 
\eeq
which determines the RS-parameter $\bar{\mu}$ as
\beq
    \bar{\mu} = \frac{\bM}{2}.
\eeq
       
In the RG-improvement with the $T$-renormalization we can only set
$\bar{\tau}(t, \rho=0)=0$, thus the resummation of terms of $O(\lambda T/\mu)$
is imcomplete. We can easily check that part of $O((\lambda T/\mu)^2)$
contributions coming from diagrams with three-point vertices at two-loop
level can not be fully absorbed through the present improvement procedure.
Such contributions can be totally
absorbed if we can set the second condition $\bar{\kappa}=0$
as in the $T_0$-renormalization case studied in the last Sec.~III.A.

With the RG improved formula  (40)-(47) in hand we can anyhow study the phase structure of the model as a result of the partial resummation of
$O(\lambda T/\mu)$ terms. 
It is surprising that all the results on the $\phi^2$-$\bM^2$ relation and
those on the corresponding EP are essentially the same as those
obtained in the $T_0$-renormalization case: the existence of the unstable small mass phase at low temperature, and the second order transition between the ordinary mass broken phase at low temperature and the symmetric phase at high
temperature.

\subsection{$T=0$ renormalization}

The application of the RG-inspired resummation procedure has been
already carried out to the one-loop EP in the $T=0$
renormalization, see Nakkagawa and Yokota in~\cite{NY2}.
Here we only would like to point out the following
fact;  If we start the perturbative calculation in the $T=0$ renormalization
scheme, then the resummation of dominant corrections of $O(\lambda (T/\mu)^2)$
should be fully accomplished through the RS-fixing condition
$\bar{\tau}(t, \rho=0)=0$, giving
\beqa
 0 &=& \ln \frac{\bM^2}{\bar{\mu}^2} - 1 +  \frac{T^2}{b \pi^2 \bM^2}
       L_1 \left( \frac{T^2}{\bM^2} \right) \\
   &\sim& 2 \ln \frac{4 \pi T}{\bar{\mu}}
         + \frac{4 \pi^2}{3} \frac{T^2}{\bM^2} - \frac{4\pi T}{\bM}.
\eeqa
Remembering $\bM^2 \sim (1/24)\blam T^2$, then Eq.(49) determines the
RS-parameter $\bar{\mu}$ as
\beq
   \bar{\mu} \sim 4 \pi T \exp (16\pi /\blam),
\eeq
showing that the remaining (un-resummed) $O(\lambda \ln (T/\bar{\mu}))$ terms
are actually large enough
to be comparable with the $O(\lambda T^2/\bM^2)$ contributions. This fact
indicates the breakdown of the resummation method
a la RG in the $T=0$ renormalization with the use of RS-fixing condition
$\bar{\tau}(t, \rho=0)=0$.

Such a trouble never happens in the $T_0$-renormalization as well as in the T-renormalization studied above in Secs.~III.A and III.B.

It is worth mentioning that, in the large-$N$ limit of the $O(N)$ symmetric
model, after setting $\bar{\tau}(t, \rho=0)=0$ there appear no
remaining (un-resummed) $O(\lambda \ln T/ \bar{\mu})$ terms, see Eqs.~(11)
and (12), assuring the absence of the above trouble in the $T=0$ renormalization.

\section{Critical Exponents}

Here we present the results of brief analysis of the critical exponents
determined from the RG-improvement one-loop EP in the $T$-renormalization,
Sec.~III.B. In this case we can calculate approximately the
critical exponents through analytic manipulation.

The definition of the critical exponents are as follows:
\begin{flushleft}
1) On the behavior at $\phi=\phi_0$ around $T \simeq T_C$:
\end{flushleft}
\beqa
  \beta  &~:~& \phi_0 \propto (T_C - T)^{\beta} \ , \\
  \gamma  &:& \left. \frac{d^2V}{d \phi^2} \right|_{\phi=\phi_0} \propto
                 | T_C - T|^{\gamma} \ , \\
  \alpha &:& V(\phi_0)-V(0) \propto |T_C-T|^{2-\alpha} \ .
\eeqa
\begin{flushleft}
2) On the behavior at $T = T_C$:
\end{flushleft}
\beq
  \delta \  : \ V(\phi)-V(\phi_0=0) \propto \phi^{\delta+1} \ \ 
  \mbox{or} \ \ \frac{dV}{d \phi} \propto \phi^{\delta} \ .
\eeq
In the above, $\phi_0$ denotes the position of the true minimum, and $T_C$
denotes the critical temperature, which are determined as
$\phi_0 \simeq \{ 54^3 T_C^4 |m^2|^3/\lambda (8 \pi)^4 \}^{1/10}$,
$T_C \simeq \sqrt{24|m_0^2|/\lambda} - 3 |m_0^2|/ 4 \pi \mu $.

Results are summarized in Table~I, in which the critical exponents
determined in the mean-field approximation, and in other works are also
given for comparison. We can see that our results show siginificant
deviations from the mean-field values, and show reasonable agreement with
the experimental results.

It should be noted that our present analysis of the critical exponents
is a very crude one, and further analysis including more prescise
detemination through numerical evaluation as well as the analysis based on
the EP determined in the $T_0$-renormalization, Sec.~III.C, are now in
progress\cite{NY3}.

\section{Discussion and comments}

In this paper we proposed a resummation method inspired by the renormalization-group improvement. By applying this resummation procedure a la RG-improvement to the one-loop effective potential in massive scalar $\phi^4$
model at $T\neq0$, we found important observations; the temperature dependent
phase transition of the model is expected to proceed through the second order transition. The critical exponents are roughly determined through analytic
manipulation, showing the siginificant deviation from the mean-field values
and the reasonable agreement with the experimental deta\cite{experi,lattice}.
This point should be made clearer with further studies on the precise determined
of critical exponents of the theory which is now in progress and the results
will be given elsewhere\cite{NY3}. With the success in the $\lambda \phi^4$
model, it is of interest to apply the present resummation method a la
RG-improvement to a more physically relevant model, such as the Abelian-Higgs
model, which is now in progress.

Discussion of the results and several comments are in order.

i) Starting the perturbative calculations with the theory renormalized at an arbitrary mass-scale $\mu$ and at an arbitrary temperature $T_0$, we can in
principle fully resum terms of $O(\lambda T/\mu)$ together with terms of
$O(\lambda (T/\mu)^2)$. The key idea is to fix the arbitrary RS-parameters so
as to make both of one-loop radiative corrections to the mass as well as to the coupling vanish. This is actually the condition which ensures the function form
of the EP to be determined up to the next-to-leading order of large correction
terms (see, the analysis in Sec.~II, just above Eq.(18)),
thus absorbing completely those terms of $O(\lambda (T/\mu)^2)$
and of $O(\lambda T/\mu)$. With the use of approximate solutions to the RGE's for
the running mass-squared we can carry out the resummation program analytically,
showing that the temperature-dependent transition between the symmetry-broken
phase and the symmetry-restored phase proceeds through the second order phase
transition. The approximation employed actually may spoil the full resummation
of terms of $O(\lambda T/\mu)$. To make this observation concerning the order
of transition in the massive scalar $\phi^4$ model more definite the
RG-improvement without use of the additional approximation, together with the
two-loop analysis are necessary, and are now under investigation.

ii) We can firstly renormalize the theory at the temperature of the environment
$T$. In this case $O(\lambda (T/\mu)^2)$-term resummation, thus the so-called
hard-thermal-loop resummation\cite{Braaten} in this model, can be simply
completed through the
$T$-renormalization itself. With the lack of freedom we can set only one
RS-fixing condition to absorb the large terms of $O(\lambda T/\mu)$, thus only
the partial resummation of these terms can be carried out. Resulting phase structure of the model is, however, essentially the same as that in the $T_0$-renormalization above. In this sense our resummation method seems to
give stable results so long as the terms of $O(\lambda T/\mu)$ are systematically
resummed.

iii) Our resummation method a la RG improvement does not work if we start
the perturbative calculation with the theory renormalized at $T=0$.
In this case the condition that may ensure
the resummation of large temperature-dependent corrections actually
generates new large terms,
thus the whole procedure of the resummation might get into trouble.

iv) As was pointed out in Sec.~III, the RG-improved EP in the simple
massive $\phi^4$ model has an unstable {\it small mass} phase at low
temperature. This unstable small mass phase also appears in the same model
at exact zero-temperature (vacuum theory), indicating its appearence
being not the artifact coming from the crudeness of the resummation of
temperature-dependent corrections terms. Though the origin of the
appearence of this unstable phase is not fully understood, it may have
a relation with the triviality of the model, which is an interesting problem
for further studies.

Finally it is worth noticing that the renormalization-group analysis of
the massive scalar $\phi^4$ model $T \neq 0$ has been done in great details
by M. van Eijck\cite{Eijck}. In his work the importance of the
finite-temperature renormalization is clearly shown. However, the
resummation of large temperature-dependent correction-terms is not the main
issue of his work, thus no idea is given on the effective
resummation-procedure of such terms, which our present work brings
into focus.

\section*{Acknowledgments}

One of us (H.~N.) thanks to Prof. G. van Weert for informing the
work by M. van Eijck.
This work was partly supported by the Special Grant-in-Aid of the
Nara University.

\newpage

\newpage

\begin{table}
\caption{Critical exponents obtained from various methods.}
\begin{tabular}{|l|c|c|c|c|} 
          & $\beta$ & $\gamma$ & $\delta$ & $\alpha$ \\ \tableline
  Our result         & 0.3  & 1.2  & 5.0  & 0.2 \\ \tableline
  mean-field         & 0.5  & 1.0  & 3.0  & 0.0 \\ \tableline
  pertur. theory & 0.5 & 1.0 & 3.0 & 0.0 \\
  (2-loop) & & & & \\ \tableline
  auxiliary mass\cite{Ogure} & ~0.385~ & ~1.37~ & 4.0 & 0.12 \\ \tableline
  nonpertur. RG\cite{BB}        & 0.35 & 1.32 & 5.0  &    \\ \tableline
  $\epsilon$-expansion\cite{eps} & 0.327 & 1.24 & ~4.79~ & 0.11 \\ \tableline
  lattice\cite{lattice} & 0.324 & 1.24 & 4.83 & 0.113 \\ \tableline
  experimental\cite{experi} & 0.325 & 1.24 &  & ~0.112~ \\
\end{tabular}
\end{table}

\newpage

%%%%%%%%%%%%%%%%%%%%%%%%%%%%%%%%%%%%%%%%%%%%%%%%%%%%%%%%%%%%%%%%%%%%
\begin{figure}\epsfxsize=8.5cm
\centerline{\epsfbox{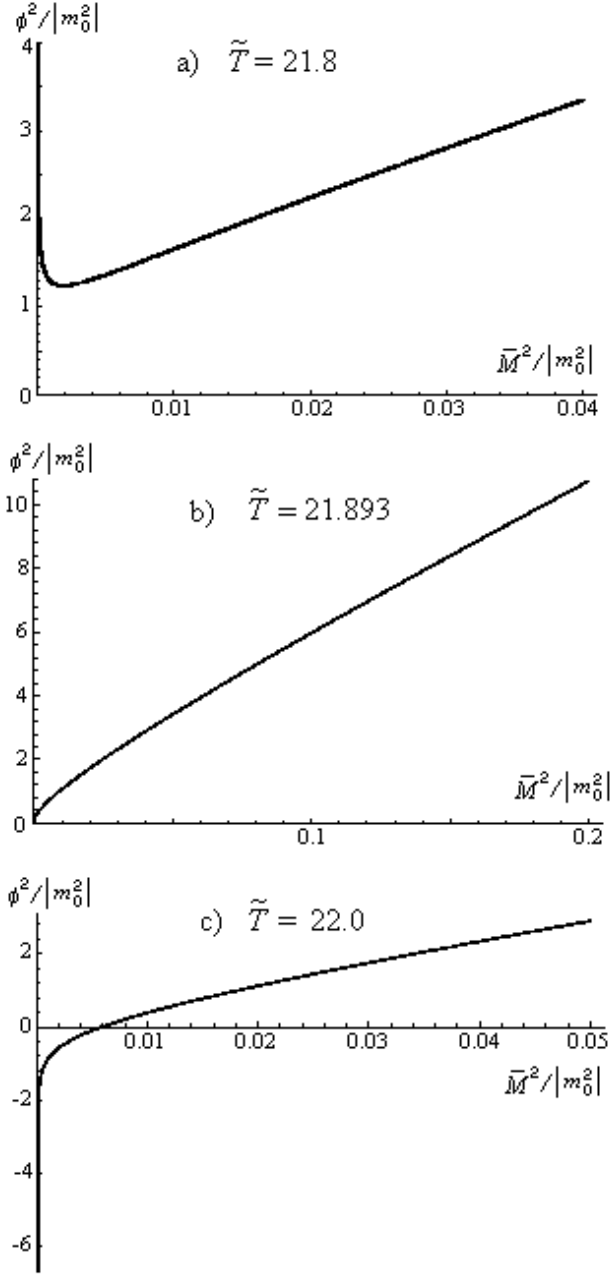}}
\vskip4mm
\caption{The $\phi^2$-$\bM^2$ relation of the massive $\phi^4$ model
      in the $T_0$-renormalization: a) $\tilde{T}_1=21.8$ ,
      b) $\tilde{T}_2=\tilde{T}_C=21.893$, and c) $\tilde{T}_3=22.0$
      with $\tilde{T} \equiv T/|m_0|$. $m_0$ is
      the renormalized mass in the vacuum theory.}
\label{fig2}
\end{figure}
%%%%%%%%%%%%%%%%%%%%%%%%%%%%%%%%%%%%%%%%%%%%%%%%%%%%%%%%%%%%%%%%%%%%
\newpage
%%%%%%%%%%%%%%%%%%%%%%%%%%%%%%%%%%%%%%%%%%%%%%%%%%%%%%%%%%%%%%%%%%%%
\begin{figure}\epsfxsize=8.5cm
\centerline{\epsfbox{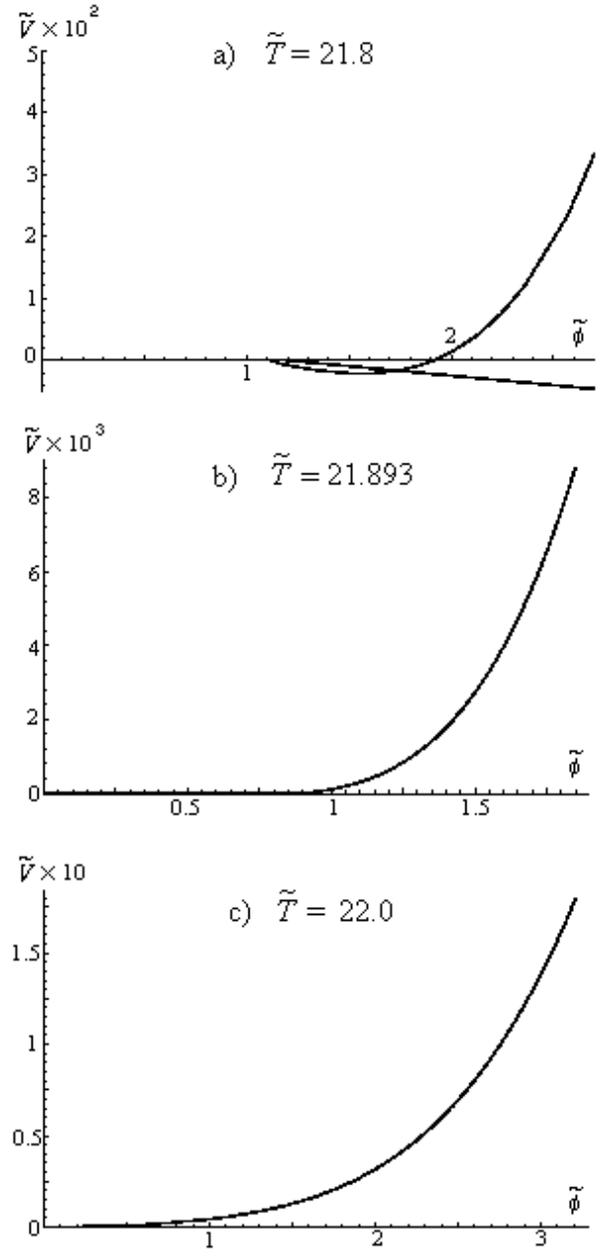}}
\vskip4mm
\caption{The RG improved effective potential of the massive $\phi^4$
      model at three temperatures in the $T_0$-renormalization: 
      a) $\tilde{T}_1=21.8$, b) $\tilde{T}_2=\tilde{T}_C=21.893$, and
      c) $\tilde{T}_3=22.0$  with $\tilde{T} \equiv T/|m_0|$.
      $\tilde{V} \equiv [ \bV_1 (\tilde{\phi}) - \bV_1 (\tilde{\phi}_{min}) ]/
      |m_0|^4$, $\tilde{\phi} \equiv \phi/|m_0|$ and $\lambda = 1/20$.}
\label{fig3}
\end{figure}
%%%%%%%%%%%%%%%%%%%%%%%%%%%%%%%%%%%%%%%%%%%%%%%%%%%%%%%%%%%%%%%%%%%%

\end{document}